\documentclass[aps,showpacs,preprintnumbers,amsmath,amssymb, 
twocolumn, tightenlines,
]{revtex4}

\usepackage[dvips]{graphicx}\usepackage{bm}
\usepackage{amsmath}
\usepackage{dcolumn}
\usepackage{bm}

\newcommand{\be}{\begin{eqnarray}}
\newcommand{\ee}{\end{eqnarray}}

\begin{document}
\title{How changing physical constants and violation of local
position invariance may occur?}

\author{ V.V. Flambaum$^1$ and E.V. Shuryak$^2$ }

\affiliation{$^1$
 School of Physics, The University of New South Wales, Sydney NSW 2052,
Australia}

\affiliation{$^2$ 
Department of Physics and Astronomy, State University of New York, 
Stony Brook NY 11794-3800, USA
}

\date{\today}

\begin{abstract}
Light scalar fields very naturally appear in modern cosmological models,
affecting such parameters of Standard Model as
 electromagnetic fine structure constant $\alpha$, dimensionless ratios of electron or quark
mass to the QCD scale , $m_{e,q}/\Lambda_{QCD}$.  Cosmological variations
of these scalar fields should occur because of drastic changes of
matter composition in Universe: the latest such event is rather recent
(redshift $z\sim 0.5$),  
 from matter to dark energy domination.   
In a two-brane model (we use as a pedagogical example) these modifications
are due to changing  distance to 
``the second brane", a massive companion
of ``our brane". Back from extra dimensions, 
 massive bodies (stars or galaxies) can also affect physical constants.
They have large scalar charge $Q_d$
 proportional to number of particles 
which produces a Coulomb-like scalar field  $\phi=Q_d/r$. This leads to a
variation of the fundamental constants proportional to the gravitational
potential, e.g. $\delta \alpha/ \alpha = k_\alpha \delta (GM/ r c^2)$.
We compare different manifestations of this effect.
 The strongest limits  $k_\alpha +0.17 k_e= (-3.5\pm 6) * 10^{-7}$ are obtained from
the measurements of  dependence of atomic frequencies on the distance
 from  Sun (the distance varies due to the ellipticity of the Earth's orbit). 
\end{abstract}

\vspace{0.1in}

\maketitle
 
\section{Introduction}

   Changing parameters of the standard model are usually associated
   with the effect of massless (strictly speaking, very light) scalar fields.
One candidate, much discussed in literature, is the so called
$dilaton$: a very special scalar which appears in string models together with a
graviton, in a massless multiplet of closed string excitations.
Other scalars naturally appear in string-theory-inspired cosmological
models, in which our Universe is a ``brane" floating in a space of larger dimensions.
The scalars  are simply brane coordinates in extra dimensions.

  At the other hand, available
  observational limits on physical constant variations at present time
  are quite strict, allowing only scalar coupling tiny in comparison with
  gravity. 
  The only relevant scalar field recently discovered -- the cosmological dark energy --
  also so far does not show  visible variations.
  So one may wander why should one
  discuss time or space-depending physics at all?

One motivation was provided by   
Damour et al \cite{Damour1,Damour:1994zq} who pointed out
that  cosmological
evolution of scalars naturally leads to their self-decoupling.    
 Damour and Polyakov have further suggested
 that  variations should happen when the scalars get excited by 
some physical change in the Universe, such as the phase transitions or other
drastic change in  the Equation of State (EoS) of the
Universe. They considered
 few of them, but since the time of their paper a new fascinating
 EoS transition has been discovered:
 from matter dominated (decelerating) era to dark energy 
dominated  (accelerating) era. It is relatively recent event, corresponding to
 cosmological redshift $z\approx 0.5$.    

  The time dependence of the disturbance related to it
  can be calculated, and it turned out \cite{Barrow,Olive} that the  self-decoupling
process is effective enough
to explain why after this transition the variation of constants is as small as observed in
laboratory
experiments at the present time, as well as at Oklo ($\sim 2$ billion
years ago or $z=0.14$) and isotopes ratios in meteorites ( $4.6$ billion years to now, $z=0.45-0$), while being
 at the same time consistent with possible observations \cite{webb1,webb2} of the
variations of the electromagnetic fine structure constant   at $z\sim 1$. 

In this paper we discuss why and how such excitation  works in some modern
cosmological models at  the same cosmological time of this transition. 
Another vast area we will address here  is similar variations of
constants in space,  near massive
bodies such as stars (Sun), pulsars, Galaxy. We will compare possible sensitivities
related with  different possible objects, point out limitations following
from some recent experiments with atomic clocks and suggest new measurements. 

 Although both authors are neither cosmologists nor string theorists, we selected
 two-brane model as our main example, for pedagogical reasons. 
 Variations of physical constants with distance to Sun and in cosmological time
have a similar nature: in both cases it is the interaction with a massive body
(Sun or  heavy ``second brane", respectively) via scalar fields. 
  We hope it will be helpful to get the message across for a non-specialized reader.

\section{Which scalars?}
 \subsection{The dilaton and its coupling functions}
 Large distance interactions in Universe are dominated
by gravity, which cannot
be screened and get cumulative effects from large bodies.
 Einstein's description -- general relativity -- uses  spin-2 metric tensor
$g_{\mu\nu}$ coupled to energy-momentum
tensor $T^{\mu\nu}\sim \delta S_m/\delta g_{\mu\nu}$ of all matter. 
 Thinking about large distance modification of gravity, people for a long time
asked why not also add a massless (or very light) spin-0 scalar field
in a similar way.

 The question is what is their nature and how are they are coupled to matter.
 One choice (originating from literature on general relativity tests) 
 is to work in the so called ``Einstein frame" and keep gravitational part of the action
 in the usual form, while  adding a scalar
 field in all other terms  via a ``coupling function" $A(\phi)$ in the metric
 \be  \tilde g_{\mu\nu}=A^2(\phi)g_{\mu\nu}^E\ee 
where $g^E$ is  the usual ``Einsteinian" metric. 
\be S ={1\over 16\pi G} \int d^4 x (g^E)^{1/2} \left[R(g^E)-2g^E_{\mu\nu} \partial_\mu \phi \partial_\nu \phi \right] \nonumber \\
+S_{matter}(\tilde g_{\mu\nu},\psi...) 
 \ee 
 If so, there is a  universal source of this field $\phi$,
\be   \Box \phi =-4\pi G T {\partial A(\phi)\over \partial \phi} \ee
containing the trace of total matter energy-momentum
  tensor $T=T_\mu^\mu$.
   This operator $T$ has a special meaning in physics, being a generator of the so called 
dilatational transformation -- a change of units of all fields.
That is why a field coupled to it is  called a dilaton field. 

Thus this field can be seen as a scale parameter: it plays the same role as e.g. a rotation
angles $\vec n$ for the rotation operator $exp(i\vec n\vec J)$ (where $\vec J$ is
angular momentum operator), or coordinate $\vec x$
in displacement operator $exp(i\vec x \vec P)$ (where $\vec P$ is momentum).
So, if $\phi(x)\neq \phi(y)$ at two points $x,y$  it means that clocks and 
meters {\em made of matter fields} (such as atoms or nuclei) 
are rescaled differently there. Note that this statement makes sense only because the purely geometrical Einsteinian metric $g^E$ of space-time 
itself was fixed independently
of $\phi(x)$, and gravitational  meters  - e.g. the Schwarzschield radius of a
black hole -- are $not$ changed.

  This field $\phi$  described so far is interacting with the total $T=\rho-3P$ of matter
 (here $P$ is the pressure). Note that it is enough 
  to violate the equivalence principle: it is say coupled to ``dust" $P=0$
  in the same way as gravity, to the mass, but not to a bag filled by
black-body radiation $\rho=3P$. Furthermore, the  energy-momentum tensor trace 
 can be expressed in terms of fundamental gauge and fermionic fields 
\be T= \sum_{i=1,2,3}{\beta_i(g) \over 2 g_i} (G_{\mu\nu}^i)^2 + \sum_f m_f\bar \psi_f \psi_f \ee
Recall that Maxwellian  energy-momentum tensor has zero trace and thus the gauge fields
of electromagnetic, weak and strong interactions ($i=1,2,3$ in the first sum)
appear in it only at the quantum level, due to the 
 so called scale anomaly induced by running coupling constant, with
$\beta_i\approx -b_i g_i^2/32\pi^2$ (in the one loop order)
being the corresponding beta functions. The weight
of say proton or nuclei include say the electromagnetic part
proportional to $\vec E^2+\vec B^2$ while the operator above contains scalar
combination $\vec E^2-\vec B^2$ with a different coefficient, they are $not$ the same.

    However,  even the string-theory dilaton is only simple at high string scale,
    where supersymmetries hold together massless supermultiplet
    which includes both the spin-2 graviton and spin-0  dilaton, together with
    a bunch of other spin 3/2,1,1/2,0 fields.  
         By the time one goes down to the scale
    at which we discuss Standard Model of particle physics 
    these symmetries should be broken by 
    complicated dynamical phenomena nobody understands.
    This may lead to a dilaton mass, and renormalize its interaction
    with different kind of matter differently.
       
    In desperation, one may thus start from the opposite extreme:
    assume  {\em different} coupling functions $A_i(\phi)$. in $all$ term of
     effective low energy Lagrangian. 
Fortunately one can still
 get rid of some of them
   by rescaling all the matter field (we already discussed
   removing scalar from gravity part above). Let us think
  about QED or QCD action, as examples. Let us start with the fermionic 
  kinetic/interaction term $\bar \psi ( i\partial_\mu -A_\mu)\gamma_\nu \psi \tilde g_{\mu\nu}$  be modified by modifying $\psi,eA_\mu$ to absorb
  $A^2(\phi)$ in it: but  now we have no freedom over the gauge field term
  \be {A_e(\phi)\over 4e^2}\int d^4x \sqrt{\tilde g} F_{\mu\nu}F_{\mu'\nu'}\tilde g^{\mu\mu'}\tilde g^{\nu\nu'} \ee because we have already fixed $A_\mu$ at the previous step. 
  The lesson is this: $\phi$ dependent ``coupling functions" can be delegated into
  the gauge field coupling constants, as Bekenstein \cite{Bekenstein}  did for QED (and extensively used in discussions of the variations such as \cite{Barrow,Olive}).

%
\begin{figure}[t]
\begin{center}
\includegraphics[scale=0.4]{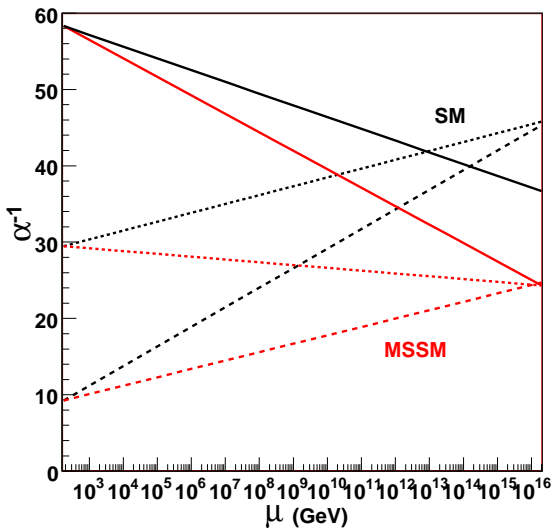}
\end{center}
\vspace{3mm}
\caption{ From \protect\cite{0701093}.
Evolution of the three gauge couplings $\alpha_1$ ($SU(5)$ 
normalized, as the dashed line), $\alpha_2$ (dotted line) and
$\alpha_3$ (solid line) in the SM (upper set of curves) and MSSM
(lower set of curves), assuming superpartner masses are all close to
the top quark mass.}
\label{fig:gut}
\end{figure}
%

  A number of authors \cite{Marciano} (see also review\cite{Uzan})
 suggested variations of em,weak and strong couplings
should be related  provided the variation comes from 
  the high energy (right in Fig.1) end.
The  strong (i=3), 
and electroweak  (i=1,2) inverse coupling constants have the following
 dependence on the scale $\mu$ and normalization point $\mu_0$:
\be \label{eqn_inv_alpha}
\alpha_i^{-1}(\mu)=\alpha_i^{-1}(\mu_0)+b_i ln(\mu/\mu_0)\ee
In the Standard Model $2\pi b_i=41/10,-19/6,-7$ and the couplings are
related as $\alpha^{-1}=(5/3)\alpha_1^{-1}+\alpha_2^{-1}$.

Fig.\ref{fig:gut} shows two popular scenarios of Grand Unification: with the standard model  as well as
for its minimal supersymmetric extension (MSSM). In the latter case one can see that 3 curves cross
at one point, believed to be a ``root" of the three branches (electromagnetic, weak and strong). One may select the unification point for $\mu_0$, and
for example, $\mu=m_Z$ is the $Z$-boson mass.

  ( String theories lead to more complicated ``trees",
which however also have a singly ``root", at a string scale $\Lambda_s$ and bare
string coupling $g_s$.)

  So, what happens when one subject the theory to a nonzero value
  of the scalar field (dilaton)?  A number of authors \cite{Marciano} (see also review\cite{Uzan})
 suggested variations of em,weak and strong couplings
should be related  provided the variation comes from 
  the high energy end.  There is quite extended discussion of what may happen,
  but basically there are two possibilities.

If one assumes that only $\alpha_{GUT}\equiv \alpha_i(\mu_0)$ varies, the eqn (\ref{eqn_inv_alpha})
gives us  the same shifts for all
inverse couplings
\be \delta \alpha_1^{-1}=\delta \alpha_2^{-1}= \delta \alpha_3^{-1}=
 \delta \alpha_{GUT}^{-1}\ee
as the whole MSSM ``tree" (Fig.1) moves up or down.
If so, the variation of the
 strong interaction constant $\alpha_3(m_z)$ is much larger than the
 variation
of the em constant $\alpha$, $\delta \alpha_3/\alpha_3=(\alpha_3/\alpha_1)\delta  \alpha_1/\alpha_1$.

Another option is the variation of the GUT scale ($\mu/\mu_0$ in eqn (\ref{eqn_inv_alpha}).  If so
 the whole ``tree" at Fig.1 moves left-right and 
quite different relations between variations of the three coupling  follows
\be \delta \alpha_1^{-1}/b_1=\delta \alpha_2^{-1}/b_2= \delta \alpha_3^{-1}/b_3 \ee
Note that now variations have different sign since the one loop coefficients
$b_i$ have different sign for 1 and 2,3. 

However, it is hard to see how this relation appears dynamically, at least  in
cosmological models on the market.  One would rather think the influence of
the ``coupling functions" depending on scalars in the matter terms of the Lagrangian would actually mean ``curving" of the lines of the ``tree" at Fig.1.  
 Another unclear issue is the
modification of lepton/quark masses, which are proportional to Higgs VEV 
and thus depend on the mechanism of electroweak symmetry breaking.

 \subsection{Introducing  cosmological two-brane model}
 
  It would be premature at this moment to subscribe to any model, of course.
   However we still find some of them more dynamically natural and possibly useful,
 at least pedagogically, to explain our intuitive ideas in relating
 distinct physics together. Field and string theory provide examples of topological
objects of different dimensions -- branes -- on which effective
 fermionic, gauge and scalar fields live. One of them may be our 4-d Universe
 with our Standard model fields,
 floating in a larger-dimensional  space\footnote{Note the similarity
 with expanding the view from the 2-d Earth surface, our natural habitat, to 4d
 Einsteinian Universe. The difference is: we can see stars and galaxies, but
  we dont know  anything definite about  these extra dimensions yet.}.   

It inspired many specific cosmological models providing some dynamical 
 realizations of 
  Brane World Cosmology (see e.g. \cite{Kiritsis,BraxDavis} for reviews).
 More specifically, our pedagogical point
is best served by the two-brane models, in which our 3-d brane 
(coordinates $\vec x$) is complemented by another 3-d brane, parallel to it but shifted 
in the 5-th dimension (we will call $y$). The Big Bang in this model is thus supplemented
not only by the influence of the ``bulk fields" living on and in between
branes
(gravitons and dilatons among them) but also by the second brane.

Without matter, symmetries of the model naturally balance all forces
between the branes -- gravity and scalar attraction is balanced by
Coulomb-like vector repulsion\footnote{There
are also some other stability conditions we dont mention, see \cite{RS2,BraxDavis}.}. It means that in this approximation
the positions of two branes $y_1,y_2$ are irrelevant: branes can ``levitate" in presence of each other, anywhere they chose. 
With extra matter added to branes this symmetry is broken and branes start
their motion (making $y_i$ a function of time $t$). If matter is not homogeneous,
as shown in Fig.2 for the left brane, the
brane  bends, 
making it a function of 3-space point $y_i(\vec x)$. In general, there appear 4-d scalar field
$y_i(\vec x,t)$,
describing the shape of the moving brane.

Unlike models mentioned above, nothing unusual happens with
physics at high energy (right) part of Fig.1: all changes are restricted to low energy
(left) end. Since gauge bosons are generated dynamically -- by open strings with
both ends on our brane -- one can calculate how much this affect
their effective Lagrangian and in principle $predict$ how they are affected by
the relative dilaton field, or $y_1,y_2$ scalars.

These two branes were originally introduced by Randall and Sundrum \cite{RS2}
to explain why gravity can be strong in general but exponentially weak
at our brane\footnote{our brane one may call  ``the brane Earth" and the other ``the brane Sun",
to enhance analogy to Copernicus/Kepler system of the world. It also explains which one
has the dominant mass.} due to warping of the space-time by gravity of the
companion brane. 

(Another fascinating aspect of the two-brane model which we will not discuss
at all is a possibility to explain the dark matter phenomenon by ascribing it to  
 the $other$ brane.)

\begin{figure}[ht!]
\begin{center}
\includegraphics[scale=0.7]{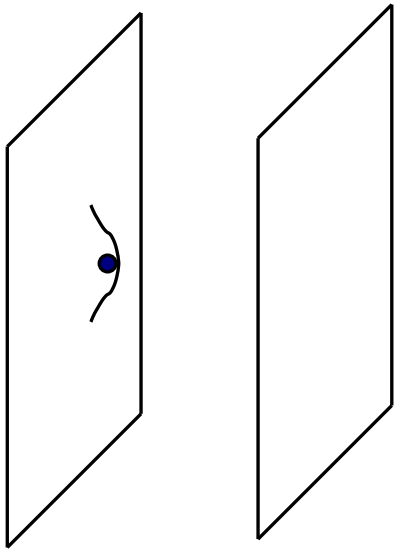}
\end{center}
\vspace{3mm}
\caption{Schematic picture of 2-brane construction. The coordinates inside
two parallel branes are called $\vec x$ and the one normal to them $y$. A massive body
(shaded circle) on a brane can cause its deformation, as shown, due to
extra attraction to the second brane. }
\label{fig:2branes}
\end{figure}

\section{The cosmological change}
\subsection{Self-decoupling mechanism}
The cosmological equations follow from Einstein's equation and are well known
 ($q=8\pi G$)
\be  -3 \ddot a /a = 2 \dot \phi ^2+8\pi G*(\rho+3P)\ee
\be 3({\dot a \over a})^2+3K/a^2=16\pi G\rho+\dot \phi^2\ee
where $a(t)$ is the scale factor, K is related to spatial curvature (below we will use K=0 flat space) and $\rho,P$ are the total energy density and pressure.
The evolution of the dilaton field is given by
\be \ddot \phi+3({\dot a \over a})\dot\phi=4\pi G \sum_A A'_A(\phi) (\rho_A-3P_A) \ee 
where the sum in the r.h.s. is over various components of matter in Universe,
with corresponding contributions to energy density and pressure\footnote{Note that simple additivity of contributions to  energy-momentum tensor is ambiguous for non-gravitationally interacting species, but
we dont expect it to be a problem for the case we aimed at,
in which one component is baryonic/dark matter and the other the dark energy.}.

    In order to understand what this equation is describing, let us 
    start
following the original Damour-Polyakov paper \cite{Damour:1994zq}
and consider a single-species Universe (no sum in the r.h.s.).
It is also useful to introduce ``logarithmic cosmic
time" $\tau=ln(a(t))+const$ we will denote derivative over it by a prime\footnote{A prime
in $A'(\phi)$ is still a derivative over $\phi$ though: we hope it will not confuse the reader.}.
The following single eqn for $\phi(\tau)$ then follows
\be \label{eqn_DP}
{2 \over 3- \phi'^2} \phi'' +(1-P/\rho)\phi'=A'(\phi)(1-3P/\rho) \ee 
which can be seen as an oscillator with exciting force (r.h.s) and the damping
(the second term in the l.h.s.) containing the EoS ratio  $P/\rho$. If its evolution
can be put in as a function of time $\tau$, this can be treated as an independent
eqn. 

For radiation-dominated Universe the r.h.s. (``excitation force")
is zero ($\rho=3P$), while l.h.s. corresponds
to damped oscillator which rapidly relaxes to rest. For other single-component
eras  the r.h.s. is not zero, but is dominated by
a single value of $P/\rho$ (e.g. zero for matter-dominated one) and a single function
$A(\phi)$: the evolution is a relaxation toward the minimum of this function,
which say happens at some $\phi=\phi_m$. 
At the minimum $A'$ is zero and all interaction of the small-amplitude dilaton with
matter vanishes.

The only time when this argument does not work is when EoS of the Universe
is changing. In a simplest case,  
there are two comparable
contributions of two types of matter. In particular, let us focus
 directly at the latest cosmological event of this change 
 in which 
components of the  energy-momentum tensor have  comparable contributions of matter
(luminous and dark) and  the dark energy (cosmological constant). A single
equation for $\phi$ (\ref{eqn_DP})
is no longer valid and one should solve all coupled equations (given above) together.
One may find explicit solutions e.g. in \cite{Barrow,Olive}, but it is easy to guess
what is happening without numerical solutions: a damped oscillator slides from
one equilibrium position 
(the minimum of coupling function to matter at $\phi=\phi_m$) to a new minimum 
of another coupling function
to the dark energy at $\phi=\phi_d$. This slide is dominated by strong damping
of cosmic oscillator, which is universal.
As a result, one is able to predict the {\em time dependence} of the
cosmological scalar field (if not the magnitude itself.).
In summary: the mechanism leads to a ladder-type evolution pattern
of the scalar field.

As shown in  \cite{Barrow,Olive}, the last slide is such that
there is actually no contradiction
between limits from Oklo and meteorite isotope compositions with
nonzero claim in quasar spectra \cite{webb1,webb2}.

There is  another version of scalar motion, in which people
want it to slide not toward a constant value now
(usually taken to be zero) but  to infinity, reproducing the so called ``tracking"
solutions. In other words, these authors  want to hit two birds with
one stone and to solve the  ``dark energy" puzzle by the same scalar. If so,
the rolling field cannot do it sufficiently quickly, 
and in this case the meteorite-based limits and
quasar data cannot be reconciled  \cite{Damour:2002mi}.

As we will discuss in the next section, however, there is no shortage
of scalar fields in modern models, and it is quite natural that different
fields would be responsible for these two phenomena.

\begin{figure}[!ht]
\hspace{5cm}
\includegraphics[width=7cm]{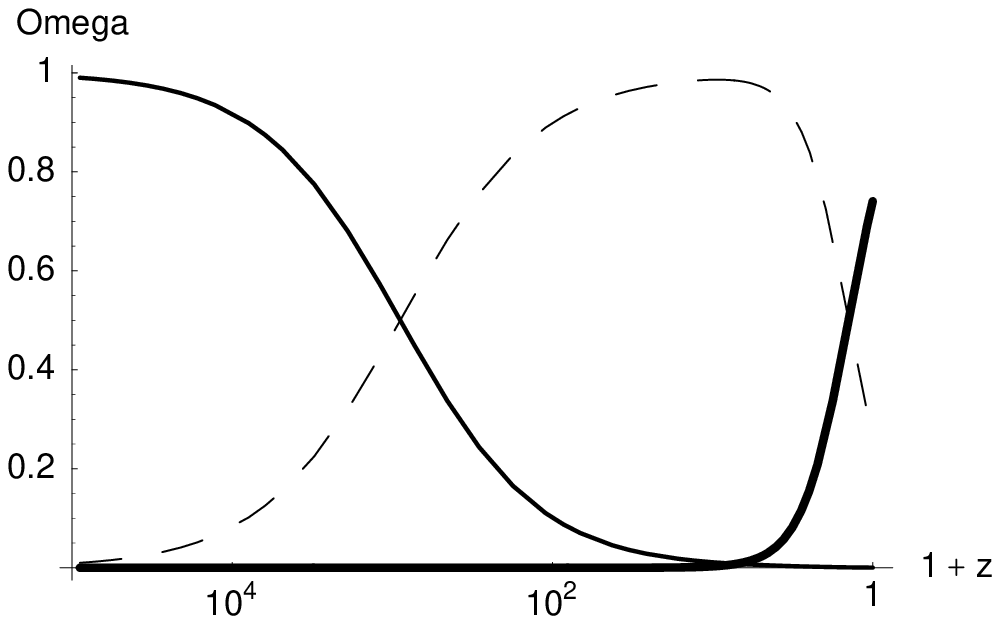}
\includegraphics[width=7cm]{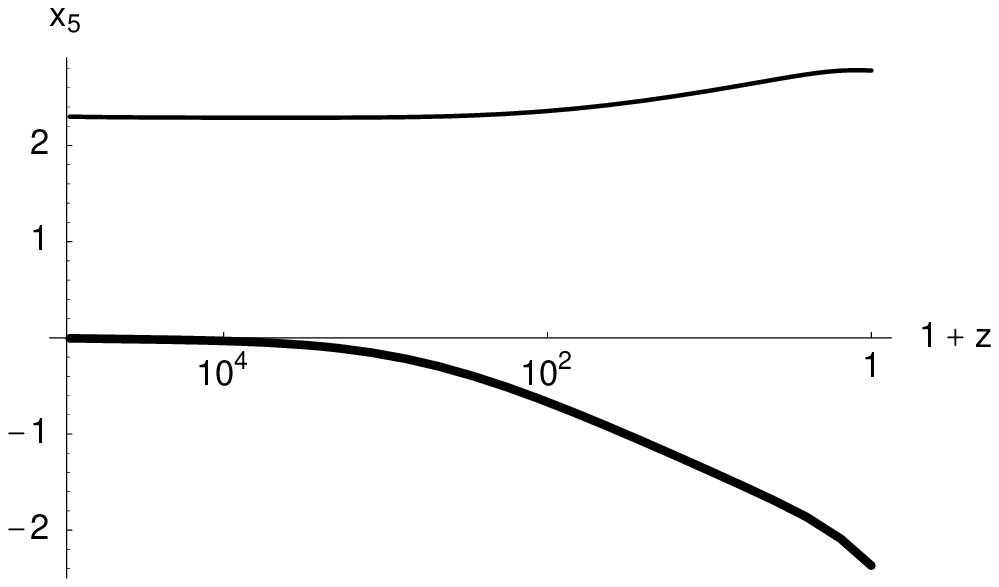}
\vspace{0.75cm}
\caption[h]{From \cite{Brax:2002nt}.
(a) Evolution of the density fraction of Universe as a
function of redshift $z$ for radiation (thin solid), matter (dashed) and the scalar fields
(thick solid). At recent time the universe is 
accelerating because of scalar dominance. Note, that in 
order to explain the values for the energy density of dark energy, 
one has to fine--tune the parameters of the theory. For these plots
 $\alpha_d=0.01$. The dark matter lives on 
the second brane, there is no matter on the first
brane. (b) The corresponding motion of the positions of two branes.
The coordinates $x_5$ was called $y$ in the text.}
\end{figure}

\subsection{Cosmic evolution of  scalars in  two-brane cosmology}

The three  scalars of the model
 are the bulk dilaton and 2 brane shapes $y_1(\vec x,t),y_2(\vec x, t)$ 
as shown in Fig. 2 : their combinations are two scalars to appear  in
the Lagrangian. The supersymmetric potential ensures ``levitation"
and defines the basic low energy Lagrangian. 
The parameters of the model include tensions (masses per area) of two branes
and bulk cosmological constant, to be tuned as explained in \cite{RS2}
for static construction. To make it dynamically moving, 
there is additional model potential expressing
the brane tension as a function of bulk dilaton scalar  which is arbitrarily selected to be exponential $U\sim exp(\alpha_d \phi)$.  As it turns out, the parameter $\alpha_d$
is restricted from above by cosmological limitations on the variations of the gravitational constant and electromagnetic fine structure  constant \cite{Brax:2002pf}.  In simple
terms, it means that the two branes should be sufficiently far apart, not to cause too much
disturbance in our world. 

As it stands, the factors depending
on scalars appear in the action 
in front of the Einstein-Hilbert $R$ term. In the language of relativity practitioners it
would be called ``Jordan frame", which is deposited by a conformal transformation
of the metric to
 the  so called ``Einstein frame"  in which there are no scalars in front of the $R$ term
 any more, with the cost of getting them in matter Lagrangian instead. 
The corresponding total low energy effective action of this model becomes of the form
\be
S_{\rm EFF} =\frac{1}{16\pi G}
\int d^4x \sqrt{-g}[ {\cal R} -  \frac{12\alpha_d^2}{1+2\alpha_d^2}
(\partial \phi)^2     \nonumber \\
 - \frac{6}{2\alpha_d^2 + 1}(\partial R)^2 ] 
-\int d^4 x\sqrt{-g} (V_{\rm eff}(\phi,R)+W_{\rm eff}(\phi,R)) \nonumber \\
+ S_m^{(1)}(\Psi_1,A^2(\phi,R)g_{\mu\nu}) + S_m^{(2)}(\Psi_2,B^2(\phi,R)g_{\mu\nu}).
\ee
where one finds two tensions (or cosmological constants) 
of the two branes $V_{\rm eff},W_{\rm eff}$ and two coupling functions
$A,B$, all of the depending on the dynamical 
scalar fields $\phi,R$ (in turn, related to the original three
scalars, the dilaton and 2 shapes). 
Writing down the evolution equations for gravity and both scalars, and going
to the cosmological setting via standard steps one gets
the following set of equations.

The so called
Friedmann equation for Hubble parameter $H=\dot a(t)/a(t)$ where $a(t)$ is
the Universe scale factor reads
\be\label{Friedmann}
H^2 = \frac{8 \pi G}{3} \left(\rho_1 + \rho_2 + V_{\rm eff}  
+ W_{\rm eff} \right) 
 \\
+ \frac{2\alpha_d^2}{1 + 2\alpha_d^2} \dot\phi^2
+ \frac{1}{1+2\alpha_d^2} \dot R^2.
\ee
The second Einstein equation is
\be\label{secEinstein}
\dot H + H^2 = -\frac{4 \pi G}{3} \left( \rho_1 + 3p_1
+ \rho_2 + 3p_2  - 2 V_{\rm eff} - 2W_{\rm eff}\right)\ \nonumber \\
-\frac{4\alpha_d^2}{1+2\alpha_d^2} \dot\phi^2
- \frac{2}{1+2\alpha_d^2} \dot R^2
\ee
The field equations for $R$ and $\phi$ read
\be
\ddot R + 3 H \dot R = - 8 \pi G \frac{1+2\alpha_d^2}{6}[
\frac{\partial V_{\rm eff}}{\partial R} +
\frac{\partial W_{\rm eff}}{\partial R}  \nonumber \\
+\alpha_R^{(1)} (\rho_1 - 3P_1) +
\alpha_R^{(2)} (\rho_2 - 3P_2) ] \label{Rcos}
\ee
\be
\ddot \phi + 3 H \dot \phi =
-8 \pi G \frac{1+2\alpha_d^2}{12 \alpha_d^2} [
\frac{\partial V_{\rm eff}}{\partial \phi} +
\frac{\partial W_{\rm eff}}{\partial \phi}  \nonumber \\
+ 
\alpha_\phi^{(1)} (\rho_1 - 3P_1) +
\alpha_\phi^{(2)} (\rho_2 - 3P_2) ].\label{Qcos}
\ee

In these expressions the matter on both branes
has energy densities and pressures $\rho_i,P_i$ entering for scalars  
is the trace of the energy--momentum
tensor for each brane's matter. 

 The coupling functions
$\alpha_\phi^{(i)}$ and $\alpha_R^{(i)}$ are defined as
\be
\alpha_\phi^{(1)} = \frac{\partial \ln A}{\partial \phi}, \hspace{0.5cm} \alpha_\phi^{(2)} = \frac{\partial \ln B}{\partial \phi};\\
\alpha_R^{(1)}=\frac{\partial \ln A}{\partial R}, \hspace{0.5cm} \alpha_R^{(2)} = 
\frac{\partial \ln B}{\partial R}.
\ee
and in the model with exponential potential  they have a particular
form, depending on the parameter $\alpha_d$ of the model. These functions describe mutual attraction of the two branes,
and thus the distance between them. 

In Fig.3(a) one finds cosmic evolution of matter composition of this model,
which is tuned to reproduce durations of radiation, matter and scalar-dominated
eras. Fig.3(b) shows the corresponding motion of the two branes.
As  one should expect from discussion in the preceding section, one finds
some sliding behavior, from equilibrium at radiation dominated era to
a gentle slide during matter era to another change recently (related
to scalar dominance and accelerated
 Universe, as now observed.

The main lesson from this is that the two-brane model leads to 
nearly the same cosmological eqns as people were discussing on more general grounds
for ad hoc scalars.  Therefore, the ideas about cosmological attractor 
leading to self decoupling of the scalars work.  In particular, in radiation-dominated
era there is no ``force" term in the r.h.s. as $\rho=3P$  and rolling to the minimum
kills evolution fast. Matter dominated period start changes in the scalars
again, as well as recent transition from matter to cosmological constant  
dominance.  

An ideal model would (i) produce the potential at the current minimum equal
to the observed cosmological constant; (ii) make relaxation to it so good that
all current and recent (Oklo, meteorites) strong limits on any 
modifications  of the constants be naturally satisfied; and maybe (iii)  produce
interesting modifications at redshift $z\sim 1$ where the matter composition
at our brane had changed the last time.  

A particular potentials studied in \cite{Brax:2002nt} numerically have not  achieved
these goals (and in fact the authors found some unexpected theoretical problems
on the way).  But the game is too recent
to give up on it, and   the settings itself is clearly sufficiently flexible  
to try many variants. 

\section{Changing physics near massive bodies}
The reason gravity is so important at large scales is that
its effect is additive. The same should be true for
 massless (or very light) scalars: its effect near large body is proportional to the
 number of particles in it.

  Unlike in the brane models, in which one can only hope to feel motion of the second
  brane via constant variations, we do see the Sun. But suppose we cannot
  (e.g. there are always clouds) and the temperature is also
  unchanged. Can one still feel that the Sun gets closer or further,
  in a periodic fashion?   By an oscillating physical constants, perhaps. 

\subsection{Scalars coupled to gravity in static  case}
 
For static star (or Galaxy) the eqn for the dilaton is
\be \Delta \phi= 4\pi G A'(\phi) T \ee
to be of course complimented by Einstein's eqn to which it is coupled
\be R_{\mu\nu}=2\partial_\mu \phi \partial_\nu \phi+8\pi G(T_{\mu\nu}-(T/2)g_{\mu\nu})\ee
It is a nonlinear eqn inside the body, which as usual for stars should be solved
starting from r=0 outward. But what one should expect to find for dilaton outside the body?

 If there is no matter there, $T=0$ at $r>R$, then the only possibility
is simply a combination of a Coulomb-like field and constant
\be \phi= Q_d/r+\phi_0\ee
Constant part is the cosmological dilaton expectation value we
discussed above: the Coulomb-like field 1/r is the subject of this
section.

But before we discuss it, let us think what would happen at constant density, say
 if one wish to include some matter such as interstellar gas or
cosmological constant, which produces small but nonzero $T$ everywhere?
If one writes $A'(\phi)=\beta \phi$ the r.h.s. is then a mass term with the
dilaton  mass
 \be m_d^2=4\pi G \beta T\ee
If T is given by the cosmological constant this mass becomes $m_d^2=6\beta \Omega_c H^2$ where $\Omega_c=0.7$ is its current balance in total density and $H$ is
current Hubble constant. Ignoring factors of the order one
one finds that this mass is only important for sizes as large as the visible
Universe itself. Thus, the first lesson thing we learned from this is that the sign
of $\beta$ should be such that there is no instability present in the Universe:
we have to be near the minimum, not maximum of the potential. 

The other lesson is that for any body the effective dilaton mass
can only gets important if its size is close to its own relativistic radius.
It is indeed precisely in this case, for which Damour and Esposito-Farese
\cite{DE-F} argued that large dilaton-induced effect is possible inside
neutron stars, provided the coupling $\beta$ happen to be specially tuned to
the star parameters making it into a kind or resonator amplifying
small cosmological dilaton field outside.

  In fact we know for sure that $outside$ neutron stars the dilaton field is small.
It was  shown in multiple works (see refs e.g. in
\cite{DE-F}) that the absence of scalar-induced corrections for binary pulsar
motion can get significant limits on that, although they are still
somewhat weaker than the limit $A'<10^{-3}$ following from solar system 
experiments related to post-Newtonian
 gravity effects.
  
  However impressive is the accuracy
  of  the available data on motion of some binary pulsars, we think that
  potentially spectroscopic studies of scalar-induced modification of
  physics constants is a more promising way to go for future studies,
  as accuracy of spectral line frequencies is still much higher.
  
For not-too-relativistic objects, like the usual stars or planets, both their total  
mass $M$ and the total dilaton charge $Q$ are simply proportional to
the number of nucleons in them, and thus the dilaton field is simply
proportional to the gravitational potential
\be \phi-\phi_0=\kappa (GM/rc^2) \ee 
and thus we expect that the fundamental constants would also depend on
the
position via the gravitational potential at the the measurement point.

\subsection{Comparison between potential and real experiments}
 Naively, one may think that the larger is the dimensionless
gravity 
potential $(GM/rc^2)$ of the object
considered, the better. However, different objects allow for quite
different accuracy. 

Let us mention few possibilities, using as a comparison parameter
the product of gravity potential divided by the tentative relative accuracy
\be P= (GM/rc^2)/(accuracy)\ee

(i) Gravity potential on Earth is changing due to ellipticity of its
orbit:
the corresponding variation $\delta (GM/rc^2)=3.3*10^{-10}$. The
accuracy of atomic clocks in laboratory conditions approaches
$10^{-16}$,
and so $P\sim 3*10^6$. However, comparing a clock on Earth and distant
satellite one may get $\delta (GM/rc^2)\sim 10^{-9}$ and $P\sim 10^7$.
The space mission was recently discussed, e.g. in the proposal 
\cite{schiller} and references therein.

(ii) Sun (or other ordinary stars) has $GM/rc^2\sim
2*10^{-7}$. Assuming accuracy $10^{-7}$ in the measurements of
atomic spectra near the surface we get $P\sim 1$.
However, a mission with modern atomic clocks sent to the Sun would have
$P\sim 10^8$ or so, see details in the proposal \cite{Maleki}.

(iii) The stars at different positions inside our (or other) Galaxy
 have gravitational potential difference of the order of $10^{-7}$,
and (like for the Sun edge) one would expect $P\sim 1$.
Clouds which give the observable absorption lines in quasar spectra
have also different gravitational potentials (relative to Earth), 
of comparable magnitude.  

(iv) White/brown dwarfs have $GM/rc^2\sim 3*10^{-4}$, and in some
cases
rather low  temperature. We thus get $P\sim 3*10^3$.

(v) Neutron stars have very large gravitational potential
$GM/rc^2\sim .1$, but high temperature and magnetic fields make
accuracy of atomic spectroscopy rather problematic, we give tentative
accuracy 1 percent. $P\sim 10$.

(vi) Black holes, in spite of its large gravitational potential, have
no scalar field outside the Shwartzschield  radius, and thus are not 
    useful for our purpose.

Accuracy of the atomic clocks is so high because they use extremely
narrow lines. At this stage, therefore, star spectroscopy seem not to
be
competitive: the situation may change if narrow lines be identified.

Now let us see what is the best  limit available today. As an example we consider recent
work \cite{clock_1} who obtained the following value for the half-year
variation
of the frequency ratio of two atomic clocks: (i) optical transitions in 
mercury ions $^{199}Hg^+$ and (ii) hyperfine splitting
in $^{133}Cs$ (the frequency standard). The limit obtained is
\be \delta ln({\omega_{Hg}\over \omega_{Cs}})=(0.7\pm 1.2)*10^{-15}\ee  
For  Cs/Hg  frequency ratio of these clocks  the dependence on the fundamental
constants
was evaluated in \cite{FT} with the result
\be \delta ln({\omega_{Hg}\over \omega_{Cs}})=-6 {\delta \alpha \over
\alpha} +0.04{\delta  (m_q/\Lambda_{QCD}) \over (m_q/\Lambda_{QCD})} -{\delta  (m_e/\Lambda_{QCD}) \over (m_e/\Lambda_{QCD})}  \ee

Another work \cite{BW} compare $H$ and $^{133}Cs$ hyperfine transitions.
The amplitude of the half-year variation  found were
\be \delta ln(\omega_{H}/\omega_{Cs}) <7*10^{-15}  \ee
The sensitivity \cite{FT}
\be \delta ln({\omega_{H}\over \omega_{Cs}})=0.83 {\delta \alpha \over
\alpha} +0.11{\delta  (m_q/\Lambda_{QCD}) \over (m_q/\Lambda_{QCD})}   \ee
There is no sensitivity to $m_e/\Lambda_{QCD}$ because
they are both hyperfine transitions.

As motivated above, we assume that scalar and gravitational
potentials are proportional to each other,  and thus introduce parameters
$k_i$ as follows

\be  {\delta \alpha \over \alpha } = k_\alpha \delta ({GM\over r c^2}) \ee
\be  {\delta (m_q/\Lambda_{QCD}) \over (m_q/\Lambda_{QCD})} = k_q \delta ({GM\over r c^2}) \ee
\be  {\delta (m_e/\Lambda_{QCD}) \over (m_e/\Lambda_{QCD}) } = {\delta (m_e/m_p) \over (m_e/m_p) } =k_e \delta ({GM\over r c^2}) \ee
where in the r.h.s. stands half-year variation of Sun's gravitational potential
on Earth. 

In such terms, the results of both experiments can be rewritten as
\be  k_\alpha +0.17 k_e= (-3.5\pm 6) * 10^{-7}\ee 
\be  | k_\alpha +  0.13 k_q | <2.5 *10^{-5} \ee

The sensitivity coefficients for other optical clocks can be found
in Refs. \cite{dzuba1999,q}, for the hyperfine clocks in Ref. \cite{FT}.
The sensitivity coefficients may be very large in  transitions between
very close levels in Dysprosium
 atom \cite{dzuba1999,Dy} ($ \delta ln(\omega_{Dy}/\omega_{Cs})\sim 10^8
\delta \alpha /\alpha$, see similar cases in \cite{nevsky}),
molecules  ($\sim 10-1000$) \cite{mol}, and optical
 UV transition in  $^{229}Th$ nucleus ($\sim 10^5$).
The sensitivity coefficients may also be very large in collision
of cold molecules near Feshbach resonance 
($\sim 10^2 -10^{12}$) \cite{cheng}.

\end{document}